# La genealogía descendente de María de Aguilar: evidencia del mestizaje colonial temprano en Costa Rica

**Bernal Morera**[1,2] **y Mauricio Meléndez Obando**[2,3]

[1]Laboratorio de Genética Evolutiva, Escuela de Ciencias Biológicas, Universidad Nacional, Heredia, Costa Rica; bernal.morera@gmail.com o bmorera@una.ac.cr
[2]Asociación de Genealogía e Historia de Costa Rica (ASOGEHI), San José, Costa Rica; melendus@yahoo.com; maomelendez@gmail.com
[3]Academia Costarricense de Ciencias Genealógicas

**ABSTRACT**

**The genealogy of Maria de Aguilar: evidence of admixture in the early Spanish Colony in Costa Rica.** During long time, historians and genealogists have interpreted that the elite that emerged during the Spanish Conquest was almost exclusively European. We reconstructed a deep matrilineal genealogy which includes recent Costa Rican ex-presidents and religious authorities back to their ancestors at the early 17th century, and compared their historic ethnic affinities with genetic mitochondrial evidence of some living descendents. The observed DNA lineage has an Amerindian ancestry. Such results point out that an Amerindian gene flow had occurred into the Spanish group during the first generations of colonial society. This conclusion do not support the current idea that the Spanish elite avoided interethnic marriages.

**KEY WORDS**

mtDNA lineages, genealogical record, ethnic history, admixture, Costa Rica.

**RESUMEN**

Por mucho tiempo historiadores y genealogistas han interpretado que la elite dominante que se formó tras la conquista española era exclusivamente europea. Se reconstruyó una genealogía matrilineal que incluye expresidentes y altos dignatarios religiosos recientes de Costa Rica, hasta sus ancestros en la sociedad Colonial temprana en el siglo XVII, y se comparó la concordancia entre las afiliaciones étnicas deducidas a partir de los registros históricos y la herencia de los linajes maternos. Se analizó el ADN mitocondrial de descendientes actuales de este linaje. El linaje ADNmt observado corresponde con una ancestría amerindia. Estos resultados corroboran que un flujo de genes amerindios tuvo lugar hacia el grupo español en las primeras generaciones de la sociedad colonial, en contraste con las ideas generalmente aceptadas de que la elite española evitaba los matrimonios exogámicos con grupos de otro origen étnico.

**PALABRAS CLAVE**

Linajes de ADNmit, genealogía, historia étnica, mezcla, Costa Rica.

El conocimiento del origen étnico de las mujeres en el periodo colonial temprano y su contribución biológica a la sociedad costarricense es un tema de muy difícil acceso desde una perspectiva documental. En la historia de Costa Rica, hay unos pocos casos de varones españoles que vinieron con esposas de su misma etnia o que, una vez consolidada la Conquista y alcanzada la estabilidad política, fueron en busca de ellas. Factores tales como la escasez de información documental acerca de los primeros colonos europeos (Meléndez Chaverri 1982), el hecho de que el grupo hegemónico durante la Colonia mostraba una tendencia a casarse con otros españoles de la misma colonia o con peninsulares (Acuña & Chavarría 1991) y que el incremento significativo en el número de individuos mestizos se volvió evidente en los documentos hasta el periodo colonial tardío, siglos XVII y XVII, (Sanabria 1977, Acuña & Chavarría 1991), han llevado a algunos académicos a suponer que la élite colonial era casi exclusivamente de ancestría española.

Una importante fuente de evidencias respecto a las mujeres fundadoras de la población costarricense proviene ahora del genoma de sus descendientes vivos –no de los documentos–, ya que sus orígenes étnicos pueden ser inferidos utilizando el enfoque de la investigación de las secuencias de ADN mitocondrial (ADNmt). Este tipo especial de material genético tiene una herencia materna estricta(es decir, son transmitidos por línea femenina exclusiva hasta un descendiente que puede estar en el presente. Por ejemplo, a los varones se les puede rastrear esa herencia genética materna aunque ellos no la transmitirán a su prole) y una gran variabilidad que lo hacen útil para realizar estudios históricos y evolutivos (Stoneking 1993). Durante los últimos años, el análisis del ADNmt ha demostrado ser un poderoso instrumento para dilucidar la historia de las poblaciones humanas. En varias otras poblaciones del continente americano que poseen una historia bien conocida de mezcla, como Brasil (Alves-Silva et al. 1999) y México (Merriwether et al. 1997), la exploración de las variantes del ADNmt ha permitido encontrar diversos niveles de contribución materna indígena, africana y europea.

Los avances tecnológicos de la genética nos han proporcionado la posibilidad de abordar el estudio de las genealogías desde una perspectiva innovadora, examinando directamente los restos óseos o analizando a los descendientes vivos a la vez que se coteja la información documental, así ha nacido la nueva ciencia de la Genealogía Molecular (Jobling 2001, Perego et al. 2005). Esta, puede ayudar a los genealogistas a descubrir relaciones familiares previamente desconocidas, a verificar los reclamos de ancestría y dar luz sobre cuestiones que han sido un rompecabezas para los historiadores familiares por años.

El surgimiento de una serie de iniciativas a gran escala para el estudio de las relaciones de parentesco entre los humanos, como el Proyecto Genográfico (National Geographic Society 1996-2010), o el equivalente de la Fundación Sorenson de Genealogía Molecular (SMGF 2010), y de una serie de compañías como la Family Tree DNA (2004) que ofrecen servicios asequibles para el análisis de ADNmt y del cromosoma Y, así como las implicaciones que tiene tal información genética en relación con el estudio de las líneas femenina y masculina, respectivamente, han contribuido enormemente a la popularización de dicha ciencia entre los genealogistas y el gran público. Un efecto es que una gran cantidad de resultados genéticos han sido publicados en la Internet por voluntad de los propios involucrados; por ejemplo, en la base Mitosearch (Family Tree DNA 2004). Ahí las personas esperan –de esta manera– localizar a sus parientes biológicos distribuidos por el mundo.

Fue precisamente en la base de datos de ADNmt de la Fundación Sorenson (SMGF 2010), donde encontramos el linaje molecular de un costarricense anónimo, cuya abuela materna fue Felicia Quirós Quirós, casada con Camilo de Mezerville (ver Madriz de Mézerville 2008). Inmediatamente identificamos que dicha persona proviene del mismo linaje matrilineal estricto de don Daniel Oduber Quirós, presidente de la República (1974-1978) y de monseñor Carlos Humberto Rodríguez Quirós, IV arzobispo de San José (nombrado en 1960), quienes están cercanamente emparentados con don José Joaquín Trejos Fernández, también presidente (1966-1970) –hijo de don Juan Trejos Quirós–, importantes personalidades de la vida nacional. Por esta razón, el resultado molecular observado adquiere una relevancia singular.

En la última década hemos venido estudiando el aporte materno al proceso de establecimiento de la población general de Costa Rica, trazando genealogías hasta las primeras generaciones coloniales y comparando los referentes documentales con los resultados genéticos de linajes mitocondriales en la población actual (Morera & Villegas 2005, Morera et al. 2010).

Desde una perspectiva etnohistórica, este tipo de estudios de casos nos abren una ventana al análisis de los problemas relacionados con el origen étnico de las mujeres fundadoras, particularmente difíciles de rastrear en los documentos, a la luz de la comparación genealógica y genética. Así, en este trabajo de minería de datos, pretendemos determinar la etnicidad de un linaje materno presidencial y examinar las implicaciones de la clasificación étnica tradicional respecto a la comprensión del origen de los costarricenses.

## METODOLOGÍA

### Análisis genealógico

Se construyó un árbol genealógico matrilineal estricto (línea uterina) a partir de información pertinente publicada en Internet (SMGF 2010) e impresa (Quirós Aguilar 1965-1966, Fernández Alfaro 1977, de la Goublaye 1982, 2000), cuyos datos relevantes fueron corroborados con los consignados en los libros de bautismos y libros de matrimonios custodiados en el Archivo Histórico Arquidiocesano Bernardo Augusto Thiel (antes Archivo Eclesiástico de la Curia Metropolitana), San José, Costa Rica. Dado que los nombres de las personas donantes de ADNmt están protegidos en la fuente original, los mantenemos en forma anónima aquí, en las dos últimas generaciones de la genealogía, para proteger la privacidad de los individuos vivos. Se presentan los nombres de las personalidades públicas relacionadas, cuyas genealogías son ampliamente conocidas.

**Análisis de los datos genéticos**

Los resultados del ADNmt del linaje que denominamos "Quirós-Quirós" fueron obtenidos a partir de la base de datos pública de la Fundación Sorenson (SMGF 2010). Para dilucidar el origen de esa secuencia del ADNmt, la comparamos con un grupo de datos de secuencias publicadas en artículos científicos que comprenden 1.119 nativos americanos, 828 africanos subsaharianos y 2.628 europeos, según se describió previamente (Morera & Villegas 2005). Para dicha comparación, se utilizó el programa ARLEQUIN (Schneider & Hudson 1996). Se confeccionó un mapa con la distribución del linaje en el continente americano.

**RESULTADOS**

La figura 1 muestra la genealogía descendente matrilineal de 13 generaciones, desde María de Aguilar [también conocida como María Calderón] hasta algunos de sus descendientes recientes. María nació hacia 1662 y fue casada con Baltasar de Segura. Este trabajo amplía y corrige algunos errores contenidos en genealogías previas (Quirós Aguilar 1965-1966, Fernández Alfaro 1977 y de la Goublaye 1982, 2000). A la fecha, no hemos encontrado registros históricos respecto a la ancestría de la madre de María de Aguilar, pero tal parece que algunos genealogistas han asumido tradicionalmente que su origen era español, pues no hacen ningún tipo de anotación sobre el particular.

El único dato que poseemos sobre la filiación de María de Aguilar (su testamento, 14 de setiembre de 1719, en Cartago) la cita como hija adoptiva del alférez José Calderón, sin mencionar el nombre de su madre.

El alférez Calderón fue "hijo legítimo" de Diego Hernández de Aguilar y María Calderón; otorgó dos testamentos, ambos en Cartago, el primero el 3 de agosto de 1705 y el segundo el 5 de setiembre de 1713. El alférez casó con Juana de las Alas (Aunque ella es consignada la mayoría de las veces como Juana de las Alas [o Juana de las Salas], en 1685 se le cita como Juana de la Cruz (en la partida matrimonial de su hija Gregoria Calderón) y en 1725 como Juana Agustina de las Alas (en el testamento de Juan de Astúa, su yerno), "hija natural" de Pedro de las Alas; ella testó mancomunadamente con su esposo en 1705 y se le cita como difunta en 1713.

En ninguno de sus testamentos, el alférez cita a su hija adoptiva, María, pero tampoco a Francisca Calderón, quien casó con el alférez Gerardo Azofeifa y había sido "criada por el alférez José Calderón". Ignoramos por qué no cita a ninguna de las dos, pero hay varias posibilidades; veamos: a) no las citó porque en realidad eran hijas de Juana de las Alas, muy posiblemente antes de haber casado con Calderón; por tanto habrían sido hijastras de Calderón, b) sí eran hijas de Calderón pero producto del "pecado" (él o la madre –o madres– de las niñas estaban casados al momento de la concepción) por lo que no podían ser reconocidas en documento público y c) realmente no eran hijas de él ni de su esposa, sino efectivamente solo hijas de crianza, tal vez hijas de algún(a) pariente(a) cercano(a).

Una curiosidad que no podemos dejar de citar, por su relación con el caso que estudiamos, es que Juana de las Alas y su padre, Pedro de las Alas, deben haber sido parientes cercanos de Francisco de las Alas, esposo de Isabel Jiménez, a quien nos referiremos en breve.

El Cuadro 1 muestra las substituciones presentes en el linaje Quirós-Quirós, en las tres regiones hipervariables del ADNmt analizadas genéticamente (HVR1, HVR2 y HVR3), comparadas con la secuencia de referencia de Cambridge (CRS) de origen europeo.

El linaje analizado (Quirós-Quirós) presenta en la región HVR1 una secuencia con cinco diferencias respecto a la referencia (CRS) en los sitios 16111T, 16223T, 16290T, 16319A y 16362C. Estas substituciones constituyen el motivo compartido por la mayoría de secuencias que pertenecen al haplogrupo A descrito por Torroni et al. (1992), el cual ha sido encontrado únicamente en indígenas americanos y asiáticos orientales. Esta secuencia con los cinco cambios diagnósticos ha sido previamente descrita en varias poblaciones amerindias de Norteamérica, Centroamérica y Suramérica, tales como los aleutianos, apaches, navajos, creeks, choctaws, ngöbés, emberás, cayapas, achés, tayacajas, tiwanakus, san martín y mapuches; y no ha sido descrita en individuos de origen europeo ni africano (Morera et al. 2010). En consecuencia, estamos en presencia de un haplotipo mitocondrial indudablemente indígena americano, el cual también ha sido encontrado en individuos mestizos de México y Brasil. La figura 2 muestra un mapa con la localización del citado linaje mitocondrial (HVRI) en los pueblos indígenas y mestizos del continente americano.

Como se puede observar (Cuadro 1), dicho linaje mitocondrial es semejante al previamente descrito a partir de varios descendientes costarricenses de Isabel Jiménez (1585-1629), hija del conquistador Alonso Jiménez y esposa de Francisco de las Alas (Morera et al. 2010). La única aparente diferencia observada entre ambos linajes consiste en un artificio metodológico de los presentes datos, que se localiza en el sitios 309.1 en la región HVR2, de la cual se carece del dato para el linaje Quirós-Quirós. En ese sitio, los descendientes de Isabel Jiménez presentan la inserción de una C, la cual no está en la referencia (CRS). La lectura de ese sitio en particular falló en el individuo del

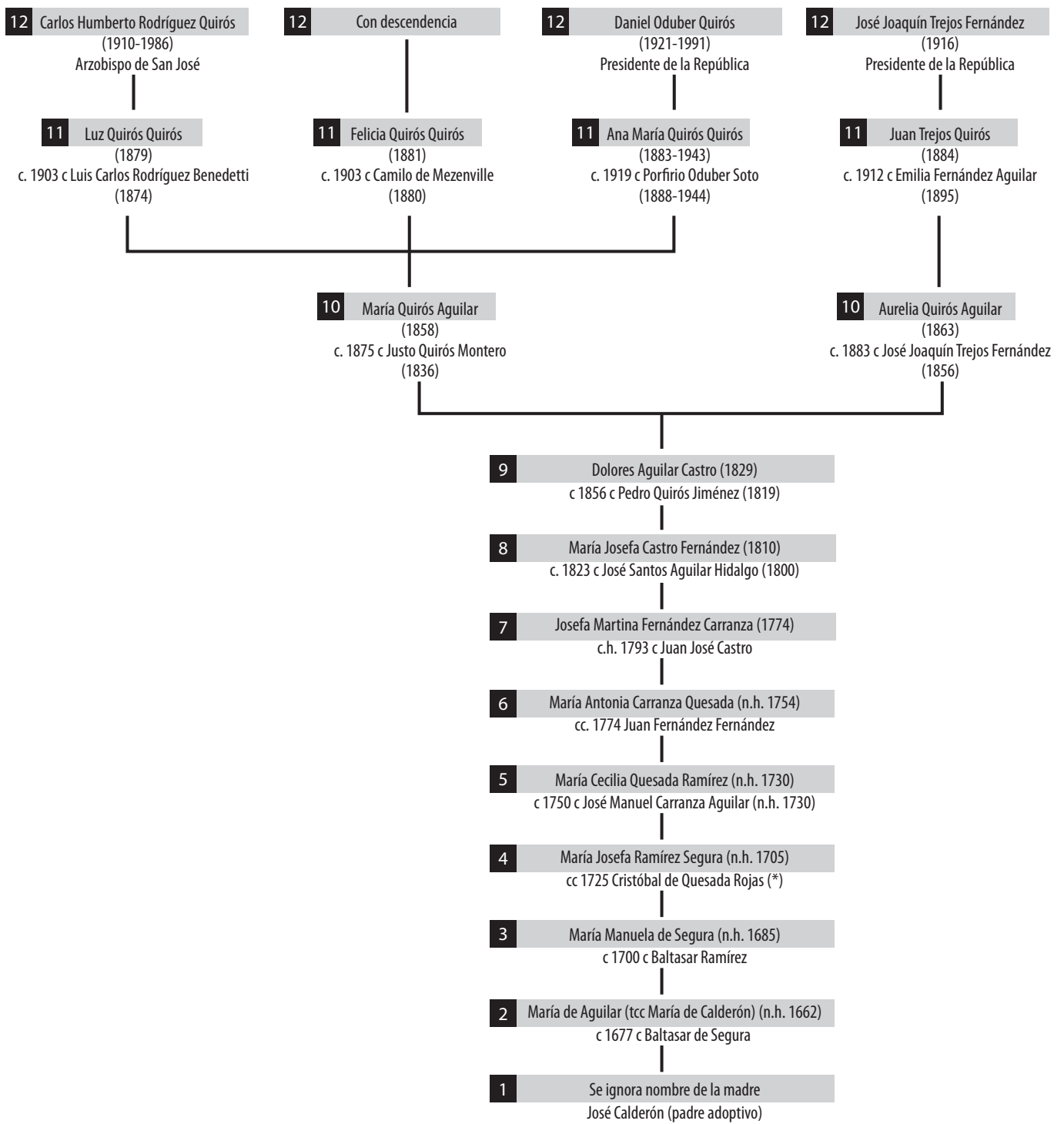

**FIG. 1.** Genealogía descendente matrilineal estricta de María de Aguilar (tcc María de Calderón) hasta algunos de sus descendientes.

(*) En el trabajo de de la Goublaye (1982, p. 299), se indican como progenitores de Cecilia Quesada Ramírez a "don Miguel de Quesada y de Roxas y doña Antonia Ramírez Segura" [sic], sin mayor explicación, cuando en realidad estos son tíos (paterno y materno, respectivamente) de Cecilia. Sin embargo, en un trabajo posterior del mismo autor (de la Goublaye 2000, p. 510), ya se cita correctamente a los padres de Cecilia Quesada Ramírez: "Cristóbal de Quesada y Rojas y Josefa de Segura", aunque no se aclara que se está rectificando un error previo. En ninguno de estos dos trabajos continúa la ascendencia de Cecilia Quesada Ramírez y en el segundo se menciona que los padres de "Josefa de Segura, olim Ramírez", son "desconocidos" (p. 214).

**CUADRO 1**
Comparación de las características genéticas del ADN mitocondrial (HVR1, HVR2, HVR3) encontradas en los descendientes vivos de María de Aguilar (linaje Quirós-Quirós), y en los descendientes de Isabel de Jiménez (~1585-1629)

**Substituciones en la Región Control del ADNmt**

| Linaje | HVR1 | | | | | HVR2 | | | | | | | | HVR3 | |
|---|---|---|---|---|---|---|---|---|---|---|---|---|---|---|---|
| | 16111 | 16223 | 16290 | 16319 | 16362 | 64 | 73 | 146 | 153 | 235 | 263 | 309.1 | 315.1 | 523 | 524 |
| CRS referencia | C | C | C | G | T | C | A | T | A | A | A | . | . | A | C |
| Isabel de Jiménez | T | T | T | A | C | T | G | C | G | G | G | C | C | d | d |
| Quirós-Quirós | T | T | T | A | C | T | G | C | G | G | G | n | C | d | d |

Notas: A: adenina, C: citocina, G: guanina, T: timina, d: deleción, (.) ausente, n: no analizado. CRS: secuencia referencia de Cambridge, HVR: región hipervariable del ADNmt.

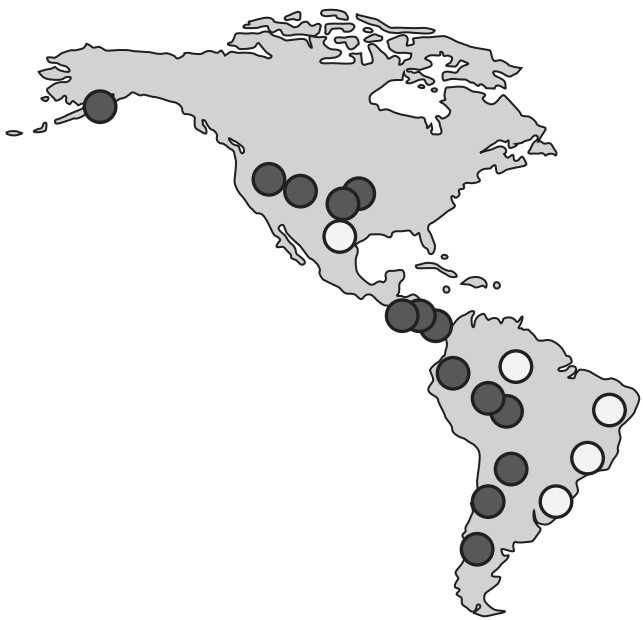

**FIG. 2.** Localización geográfica de los pueblos indígenas (en gris oscuro) y mestizos (en blanco) actuales en los cuales viven individuos emparentados matrilinealmente que poseen exactamente el mismo linaje mitocondrial (haplotipo de la HVRI) encontrado en los descendientes costarricenses de María de Aguilar y de Isabel de Jiménez.

linaje Quirós-Quirós de modo que tenemos un vacío en la información por lo que no sabemos cuál es la base (letra) presente en dicha posición. Y por tanto no sabemos si existe o no una diferencia genética entre dichos linajes en ese único sitio. La forma de solventar tal deficiencia sería realizando en el futuro la prueba genética a más individuos del linaje. Lo que sí sabemos a ciencia cierta es que en el resto de la longitud de la secuencia informativa –de cientos de pares de bases–, ambos linajes son exactamente iguales. Estos resultados genéticos indican que ambos linajes están emparentados, compartiendo ambos una antepasada común.

## DISCUSIÓN

### Origen étnico del linaje

Los resultados genético-genealógicos del linaje Quirós-Quirós, restringidos más a la existencia de descendientes vivos que a los documentos, nos ofrecen información reveladora respecto a algunos de sus ancestros maternos. De la inquebrantable sucesión de madres e hijas que se remonta en el tiempo, podemos distinguir cuatro mujeres de carne y hueso que vivieron en diferentes épocas, y conocer algunos detalles de sus vidas.

Primera, María de Aguilar. En ausencia de documentos, conocemos ahora que su origen étnico era mestizo.

Considerando las fechas, ella sería muy posiblemente lo que luego llamaron despectivamente "cuarterona de india". Estaba, empero, lo suficientemente integrada a la élite colonial, como para que sus retoños se unieran por matrimonio con varones españoles, criollos y peninsulares. Con el tiempo algunos de sus descendientes directos llegaron a ser presidentes o altos jerarcas eclesiásticos de Costa Rica (Quirós Aguilar 1965-1966, de la Goublaye 2000).

A pesar de que no se han encontrado documentos que lo ratifiquen, los datos biológicos indican que María de Aguilar e Isabel de Jiménez estuvieron ineludiblemente emparentadas por línea materna. Existe la posibilidad de que María fuera sobrina-nieta de Isabel. Aunque también existe la posibilidad de que el parentesco sea un poco más lejano y su antepasada común femenina se remonte unas generaciones atrás a tiempos precolombinos. El ulterior análisis del ADNmt de más miembros de este linaje nos permitirá discernir sobre estas dos alternativas.

Debemos anotar, como dato curioso, que Manuela de Segura, hija de María de Aguilar y el sargento Baltasar de Segura, habría sido parienta de su marido, el alférez Baltasar Ramírez, pues este fue hijo del teniente Salvador Ramírez y Magdalena Milanés; a su vez, Magdalena Milanés fue hija de Pablo Milanés del Castillo y Catalina de las Alas, esta última, hija de la precitada Isabel Jiménez. Por tanto, Manuela de Segura y Baltasar Ramírez compartían una misma antepasada por línea mitocondrial.

Segunda, la madre o más probablemente la abuela o bisabuela de María, debió ser una mujer indígena americana, que vivió durante la Conquista y el inicio de la Colonia española. El haplogrupo A2, al cual pertenece este linaje es el más frecuente en los indígenas costarricenses (Santos et al. 1994). Esta dama de nombre desconocido supo sobrevivir y adaptarse al cambio en su mundo, como lo demuestran el número y éxito biológico y social de sus descendientes.

A la tercera mujer, podemos acceder gracias a la amplia e inusual distribución geográfica de este haplotipo –perteneciente al haplogrupo A2, el cual nos da cuenta de que estamos en presencia de un linaje fundador americano (Morera et al. 2010). Sus descendientes pueden sentirse muy orgullosos al saber que una de sus "abuelas" estuvo efectivamente entre los verdaderos primeros descubridores, conquistadores y pobladores del continente americano, quienes llegaron al Nuevo Mundo, unos 20.000 años antes que los vikingos y más aun que los españoles, atravesando las tierras de Beringia, entonces sobre el nivel del mar (National Geographic Society 1996-2010). Sus descendientes directos continúan viviendo hoy día, desde las islas Aleutianas hasta el centro sur de Chile.

Los datos biológicos nos remiten a una cuarta persona relacionada con el origen del haplogrupo A. Vivió hace unos 50.000 años probablemente en las altiplanicies de Asia Central entre el Mar Caspio y el lago Baikal. Los grupos humanos que se movían al este llevaron el haplogrupo A con ellos y se dispersaron en varias áreas de Asia Oriental. Sus descendientes se encuentran con alguna frecuencia en el sureste asiático, Asia Central, Siberia y en las poblaciones coreana, china y japonesa. Hoy en día, un subgrupo de este, el haplogrupo A2 es uno de los cinco linajes mitocondriales encontrados en los indígenas americanos y se encuentra tanto en Norteamérica como América Central y del Sur (National Geographic Society 1996-2010).

**Implicaciones etnohistóricas respecto a la fundación del pueblo costarricense**

La población general de Costa Rica es el resultado de una mezcla trihíbrida, con proporciones globales de genes de origen europeo, amerindio y africano de 61.04%, 29.91% y 9.05%, respectivamente (Morera & Barrantes 1995, Barrantes & Morera 1999, Morera et al. 2003). Sin embargo, muchos aspectos del proceso de mezcla no se comprenden claramente aún; por ejemplo, el fenómeno del apareamiento direccional, donde los hombres europeos y las mujeres indígenas o africanas habrían contribuido en forma desproporcionada a la formación de las poblaciones actuales. Es así como nuestro estudio puntual del linaje Quirós-Quirós corrobora que la mezcla español-indígena comenzó en Costa Rica desde las primeras generaciones de la sociedad hispanoparlante (Morera et al. 2010), y no hasta el siglo XVIII como se había sugerido previamente (Acuña & Chavarría 1991). El "orgullo de su ancestría española" fue tradicionalmente preservada como un valor en algunas familias de la élite, durante un periodo de casi 300 años (Meléndez Obando 1999), y no obstante en la mayoría de los casos estudiados (5 de 6 a la fecha), se demuestra el ascendiente indígena, aun en este segmento de la población en el cual podríamos esperar menos influencia de la población amerindia.

El historiador Carlos Meléndez Chaverri (1982) en su libro Conquistadores y pobladores: orígenes histórico-sociales de los costarricenses, ofreció una notable reconstrucción la fundación del pueblo costarricense, que sin embargo nos trae a la memoria aquellas descripciones de la Atenas clásica, basadas casi exclusivamente en los ciudadanos varones. Con muy pocas menciones a los constructores de los magníficos templos y estatuas: los esclavos, y aún menos referencias a las mujeres. Entendemos que, al restringir Meléndez Chaverri (1982) el supradicho trabajo al siglo XVI (o sea a un lapso muy pequeño de tiempo que incluye la Conquista y solamente el inicio de la Colonia) y a los grupos conquistadores, sesgó el resultado de su análisis y eso le impidió una valoración global del papel que desempeñaron los otros integrantes del proceso, como los indígenas, los subsecuentes europeos y los africanos, igualmente fundadores. Por otra parte, Meléndez Chaverri (1982) nos explica que las escasas menciones a las mujeres, se deben a la dificultad que tuvo para obtener documentación referente al papel desempeñado por ellas durante todo el proceso.

Meléndez Obando (2004) retomó el tema, solventando la deficiencia antes mencionada con información documental de carácter genealógico, logró así rescatar el nombre de 59 esposas o "compañeras" de los ahora 102 conquistadores o pobladores conocidos –de igual forma limitado artificialmente al periodo 1561-1599– (véase Cuadro 2). Tal revisión permite concluir que estrictamente solo nos consta que Da. Mayor de Benavides (esposa Juan de Solano), Da. Inés Ruiz de Villatoro (esposa de Alonso del Cubillo), Da. Sabina de Artieda (esposa de Juan de Peñaranda) y Da. Juana Chacón (segunda esposa de Francisco Pavón) eran españolas. Además, se citan con el tratamiento de "Doña": Ana (esposa Antonio de Carvajal), Inés de Ampuero (¿primera? esposa de Francisco Pavón), Andrea Vázquez de Coronado (esposa de Diego Peláez

**CUADRO 2**
Conquistadores y pobladores de Costa Rica (1561-1599)

| | Conquistador o poblador | Esposa o "compañera" |
|---|---|---|
| 1 | Álvaro de Acuña (1535) esp VC | Catalina de Acuña e |
| 2 | Pero Afán de Ribera* (1492) esp | Da. Petronila de Paz e |
| 3 | Diego de Aguilar (1548) VR | Catalina Palacios e |
| 4 | Hernando de Aguilar (1545) ART | |
| 5 | Pero Alonso de las Alas (1536) C | María de Guido nic e hc |
| 6 | Cristóbal de Alfaro (1540) P | Catalina Gutiérrez Jaramillo e hc |
| 7 | Juan Alonso** (1534) C | |
| 8 | Antonio Álvarez Pereira (1530) por C | Da. Inés [¿india?] |
| 9 | Francisco de Arrieta (1575) F | Catalina Gómez e |
| 10 | Diego de Artieda Chirinos* esp | María de Céspedes e |
| 11 | Juan Barbosa C | María Verdugo (Nicaragua) e |
| 12 | Martín Beleño** F | |
| 13 | Román Benito (1530) C | Juana Gómez e |
| 14 | Alonso de Bonilla (1556) nic VC | Ana López de Ortega e hc |
| 15 | Francisco de Bonilla** VC | |
| 16 | Ambrosio de Brenes (1569) F | María de Espinosa e hc |
| 17 | Jerónimo Busto de Villegas C | |
| 18 | Juan Cabral** ART? | Catalina Gutiérrez Jaramillo e hc |
| 19 | Miguel Calvo esp F | Mariana Chinchilla hond e |
| 20 | Antonio de Carvajal VC | Da. Ana e |
| 21 | Diego del Casar Escalante P | |
| 22 | Luis Cascante de Rojas F | Juana Solano e hc |
| 23 | Ignacio Cota (1530) esp C | |
| 24 | Alonso del Cubillo (1540) esp A | Da. Isabel Ruiz de Villatoro esp e |
| 25 | Cristóbal de Chaves (1569) F | María de Alfaro e hc |
| 26 | Gaspar de Chinchilla (1540) F | Catalina Palacios e hc |
| 27 | Gaspar Delgado (1540) P | María del Castillo e hc |
| 28 | Pedro Díaz de Loría (1538)   VC | Magdalena Ojeda  e |
| 29 | Luis Díaz de Trejo**  C | |
| 30 | Pedro Enríquez de Cadórniga** esp  P | |
| 31 | Francisco de Estrada** (1527)   C | |
| 32 | Luis de Esquivel Añasco   F | |
| 33 | Alonso Fajardo**  VC | |
| 34 | Hernando Farfán   C | Antonia de Trujillo  e  y Catalina Rueda  e  nc |
| 35 | Jerónimo Felipe (1568) esp  F | María de Ortega  e  hc |
| 36 | Leandro de Figueroa (1566)  esp  F | Inés Solano  e  hc |
| 37 | Pedro de Flores (1554)   A | Isabel Juárez  e  hc |
| 38 | Francisco de Fonseca  VC | Catalina Hernández  e |
| 39 | Juan García   P | Juana Carrillo  e |
| 40 | Pedro García Carrasco (1542)  VC | Jerónima de Avila  e |
| 41 | Francisco Ginovés [Ferreto] (1512)  C | |
| 42 | Alonso Gómez Macotela   F | |
| 43 | Baltasar González (1532)**  VC | |
| 44 | Alonso de Guido (1543) | Isabel Núñez  e |
| 45 | Alonso Gutiérrez de Sibaja (1541) gua  C | María Alvarez de Oviedo nic  e |
| 46 | Domingo Hernández (1536)   C | |
| 47 | Antonio Hernández Camelo   VC | |
| 48 | Antonio de Herrera**  VC | |
| 49 | Francisco Hidalgo**  A | Inés Pérez  e |
| 50 | Alonso Jáimez esp  F | |
| 51 | Diego Jáimez**  F | |
| 52 | Gómez Jaramillo   P | Magdalena Gutiérrez  e |
| 53 | Alonso Jiménez (1544) esp  VC | |
| 54 | Domingo Jiménez (1534)   VR | Gracia [india] |
| 55 | Juan Jiménez  esp  VC | |
| 56 | Francisco Lobo de Gamaza** (1534)  VC | |
| 57 | Fernando López de Azcuña   F | |
| 58 | Juan López Cerrato de Sotomayor*** esp  A | |
| 59 | Juan López de Ortega (1546)   P | Catalina de Ortega  e  y María López [india] |
| 60 | Sebastián López de Quesada   P | |
| 61 | Cristóbal de Madrigal  VC | Elvira Gómez  e |
| 62 | Francisco Magariño  VC | |
| 63 | Pedro Luis Medina Cueto   F | Isabel de Carvajal  e  hc |
| 64 | Esteban de Mena (1532)   C | |
| 65 | Vicencio Milanés (1522)**  it | |
| 66 | Martín de Miranda**   VC | Inés Hidalgo  e |
| 67 | Felipe Monge (1565) esp  F | Francisca López  e  hc  nc |
| 68 | Morales  F | |
| 69 | Francisco de Ocampo Golfín (1570) esp  F | Inés de Benavides  e  hc  y Catalina Tuia [india] |
| 70 | Francisco Ochoa (1540)   ART | |
| 71 | Juan Ordóñez del Castillo (1535)  C | |
| 72 | Matías de Palacios (1550)   P | Luisa Hernández  e  hc |
| 73 | Francisco Pavón** ART | ¿Da. Inés de Ampuero e?, Da. Juana Chacón e |
| 74 | Diego Peláez de Berríos (1565) F | Da. Andrea Vásquez de Coronado e nc |
| 75 | Juan de Peñaranda (1535) esp ART | Sabina de Artieda esp e |
| 76 | Antonio de Peralta esp VC | Juana del Moral e |
| 77 | Gaspar Pereira Cardoso (1570) por F | Isabel de Acuña e hc |
| 78 | Juan Pereira VC? | |
| 79 | Juan Pérez (1522) F | Francisca de Avila e Inés [india] |
| 80 | Alonso Pérez Farfán (1527) esp C | Marina de Anangas esp [no vino a CR] |
| 81 | Pedro de la Portilla (1547) A | Ana Gómez e hc |
| 82 | Agustín Félix de Prendas** (1558) ART | Beatriz Fernández y Da. Tomasina de Lerma |
| 83 | Diego de Quesada (1552) esp ART | Francisca Gutiérrez de Sibaja mst e hc |
| 84 | Diego Quintero** C | |
| 85 | Francisco Ramiro Corajo P | Francisca de Zúñiga e |
| 86 | Jerónimo de Retes (1560) esp F | María de Ortega e hc |
| 87 | Pedro de Ribero y Escobar (1538) VC | Catalina de Vega e |
| 88 | Nicolás Rodas (1567)** F | |
| 89 | Francisco Rodríguez** P | |
| 90 | Gaspar Rodríguez (1550) A | Inés Rodríguez e y María Ramírez e |
| 91 | Juan Rodríguez Calderón** (1550) P | |
| 92 | Diego Rodríguez Chacón VC | |
| 93 | Domingo Rodríguez Portugués F | |
| 94 | Bartolomé Sánchez (1555) P | Da. Inés Alvarez Pereira mst e hc |
| 95 | Miguel Sánchez de Guido (1528) esp C | Leonor de Mendoza e hc |
| 96 | Juan Solano (1538) esp C | Da. Mayor de Benavides esp e |
| 97 | Juan Suazo1 | Isabel Rodríguez e |
| 98 | Salvador de Torres (1576) esp F | Inés Pérez Farfán e hc |
| 99 | Diego de Trejo** (1538) C | |
| 100 | Jerónimo Vanegas (1551) mex VR | Teresa Fernández [no vino a CR] e |
| 101 | Blas de Vera Bustamante** VC | |
| 102 | Diego de Zúñiga F | |

Nota: Se consideran fundadores aquellos que ingresaron antes de 1599 y dejaron descendencia que vivía aún en la primera cuarta parte del siglo XVII. Como hemos seguido el modelo de Meléndez Chaverri (1982) -base de este listado, habrá un sesgo pues él no considera a las mujeres como fundadoras, solo bajo el concepto grupal de familia (aunque no necesariamente la hayan integrado como tal). Asimismo, si se considerara fundador a cada español que arribó en ese periodo, la cifra subiría considerablemente pues algunos llegaron con hijos varones y mujeres, como Alonso del Cubillo y Juan de Peñaranda, cuyos hijos no están en la presente lista.

(*) Aunque Meléndez Chaverri (1982) los incluye como fundadores, no los considero tales pues no hay prueba documental, ni indicios, de que hayan dejado sucesión en Costa Rica.

(**) Caso dudoso porque no hay pruebas contundentes de que sean fundadores.

(***) La filiación dada por Meléndez Chaverri (1982) es totalmente equivocada. Juan López Cerrato de Sotomayor fue hijo legítimo de Alonso Fernández de Córdoba Sotomayor y Da. Inés Cerrato [esta última hija legítima del Dr. Juan López Cerrato y María de Contreras, naturales de España, quienes también tuvieron un hijo nombrado Juan López Cerrato, vecino de Granada, Nicaragua]. Tampoco lo consideramos fundador.

(1) Se ignora cuándo ingresó a Costa Rica, pero sabemos que fue antes de 1599. Se le cita como poblador antiguo de la provincia.

Abreviaturas: A: entró con Anguciana de Gamboa; ART: entró con Artieda Chirinos; C: entró con Cavallón; CR: Costa Rica; Da.: doña; e: esposa; esp: natural de España; F: entró entre 1590 y 1599; gua: natural de Guatemala; hc: hija de conquistador; hond: Honduras; mex: natural de México; mst: mestiza; nc: nieta de conquistador; nic: natural de Nicaragua; P: entró con Perafán de Ribera; por: portugués; VC: entró con Vásquez de Coronado; VR: entró con Venegas de los Ríos;

FUENTE: Meléndez Obando, M.O. 2004.

de Berríos) y Tomasina de Lerma (esposa de Agustín Félix de Prendas), lo cual sugiere un origen español –documentado– para ocho de las mujeres fundadoras. Se infiere de esto, que serían de etnia española –con bastante seguridad, tan solo– un 13.56% (8 de 59) de ellas.

Recientemente, de la Goublaye de Ménorval (2008) abordó de nuevo el tema, desde una perspectiva de caso focal, restringido a sus ancestros personales durante más o menos el mismo periodo de tiempo. Tal esfuerzo no es desdeñable, ya que su número es bastante grande y quizás sea una muestra estadísticamente significativa, considerando que Meléndez Chaverri (1982) sugirió 86 familias fundadoras y Meléndez Obando (2004) lo amplía a poco más de 100. Y como quiera que sea, estos son también ancestros de la inmensa mayoría de los costarricenses con raíces en el Valle Central. De acuerdo con la clasificación ofrecida por de la Goublaye de Ménorval, se concluye que un 76.92% (49 de 65) de las mujeres fundadoras de familias del siglo XVI tenían un origen étnico ibérico, un 16.92% (n= 11) indígena o mestizo, y un 6.16 % (n= 4) origen desconocido. Como comentaremos adelante, estos valores podrían estar sesgados.

Aún cuando los criterios de clasificación y cuantificación son ligeramente distintos, es notable la diferencia entre estos dos últimos estudios, respecto a las estimaciones del aporte europeo materno al origen de la población de Costa Rica. Hecho que amerita ser analizado.

Es en este contexto que los estudios combinados de genealogía y genética (Morera et al. 2010) –como el presente– adquieren relevancia, ya que nos permiten explorar asimismo las nociones de etnicidad que se pueden interpretar a partir de los documentos con la realidad biológica de aquellas personas de la época colonial, inferida a partir de sus descendientes. Así, por ejemplo, si un genealogista hubiera clasificado la afiliación étnica de María de Aguilar, siguiendo los criterios apriorísticos de los trabajos históricos y genealógicos tradicionales (p.e. nombre hispano, casada con un español, habitante de una ciudad colonial, etc.) y en ausencia de documentos explícitos respecto a su origen (p.e. salida de España, uso en vida del distintivo de "doña"), probablemente habría supuesto que ella era española. Sin embargo, tal hipótesis ha resultado ser errónea. Si bien doña María no está incluida en la lista analizada por de la Goublaye de Ménorval (2008) pues nació en el siglo XVII, la ejemplificación es pertinente, porque otras de las damas coloniales clasificadas ahí como de origen ibérico, biológicamente no lo eran. Pero este es tema de otro trabajo.

Es probable que tanto como un 80% de los casos en que se tenga el nombre hispano de una matrona colonial, en realidad se esté en presencia de un linaje mitocondrial no europeo, y por tanto mestizo. En conclusión, asumir a priori una afiliación étnica ibérica para las mujeres fundadoras de la población costarricense es muy probablemente un error.

## REFERENCIAS